\documentclass[aip,pop,reprint,floatfix]{revtex4-1}

\usepackage{graphicx}
\usepackage{appendix}

\usepackage[utf8]{inputenc}
\usepackage[T1]{fontenc}
\usepackage{mathtools}
\usepackage{cancel}
\usepackage{adjustbox}
\usepackage{multirow}

\newcommand{\beq}{\begin{equation}}
\newcommand{\eeq}{\end{equation}}

\newcommand{\vu}{\boldsymbol{u}}
\newcommand{\vB}{\boldsymbol{B}}
\newcommand{\vb}{\boldsymbol{b}}

\newcommand{\vA}{v_{\mathrm{A}}}

\newcommand{\dd}{\partial}

\newcommand{\ddt}{\dd_t}
\newcommand{\ddx}{\dd_x}
\newcommand{\ddtau}{\dd_\tau}

\newcommand{\Dt}{\mathrm{d}_{t}}
\newcommand{\kpar}{k_\parallel}

\newcommand{\kperp}{k_\perp}

\newcommand{\tn}{\tilde{n}}

\newcommand{\avg}[1]{\overline{#1}}
\newcommand{\dpar}{d_\parallel}
\newcommand{\hmu}{\hat{\mu}}
\newcommand{\ddze}{\dd_\zeta}
\newcommand{\dndxisq}{[\mathrm{d}n_0/\mathrm{d}\xi]^2}
\newcommand{\ddxi}{\dd_\xi}

\newcommand{\markup}[1]{{#1}}

\draft 
\begin{document}

\title{Nonlinear dynamics of large-amplitude, small-scale Alfv\'en waves} 

\author{Alfred Mallet}
\email[Email address for correspondence: ]{alfred.mallet@berkeley.edu}
\affiliation{Space Sciences Laboratory, University of California, Berkeley, CA 94720, USA}
\date{\today}

\begin{abstract}
We study large-amplitude, very oblique Alfv\'en waves at low $\beta$, with small gradient length scales, comparable to the ion inertial scale $d_i$. Such waves have large density fluctuations, and slight dispersion from finite-frequency and finite ion sound radius effects. We derive a weakly nonlinear evolution equation governing the behaviour of the waves in one dimension, and categorize the different solitons appearing in different regimes: the regular solitons involve full rotations of the transverse magnetic field, similar to modified Korteweg-de Vries (mKdV) solitons (our nonlinear equation reduces to the mKdV equation in the long-wavelength limit). However, for sufficiently small soliton widths, some become singular, small-amplitude solitons with density discontinuities, and are thus expected to become strongly dissipative in a real plasma. These solutions may be useful in explaining some aspects of the sharp, ion-scale magnetic field rotations (switchbacks) observed in the near-Sun solar wind by Parker Solar Probe.
\end{abstract}


\maketitle 

\section{Introduction}
The Alfv\'en wave (AW) is unique among the MHD waves: within the MHD framework, any (even three-dimensional) configuration whose velocity and magnetic field fluctuations satisfy
\beq
\delta \vu = \pm \delta \vB /\sqrt{4\pi\rho},\label{eq:alfpol}
\eeq
respectively, and also maintains a constant density, pressure, and magnetic field strength
\beq
n = \text{const}, \quad P = \text{const}, \quad B^2 = \text{const}.\label{eq:stconst}
\eeq
propagates \textit{without steepening}, regardless of its amplitude, at the Alfv\'en velocity
\beq
\boldsymbol{v}_A = \mp\avg{\vB}/\sqrt{4\pi n m_i}\label{eq:vadef}
\eeq
where $\avg{\vB}$ is the mean magnetic field and $n$ is the number density of the ions with mass $m_i$. This exact nonlinear solution is the MHD large-amplitude AW\cite{goldstein1974}. At long wavelengths this solution survives relatively unscathed even in a collisionless plasma: for wavelengths much larger than the ion scales, AW are undamped\cite{barnes1971,barnes1974,schektome2009}, unlike slow and fast waves. 

Large-scale propagating fluctuations that are not in the Alfv\'enic state generically steepen into shocks and/or dissipate due to wave-particle interactions. This may explain why AW are ubiquitous in the solar wind\citep{belcher1971}, especially close to the sun and in the corona, currently being explored by NASA's Parker Solar Probe (PSP)\citep{bale2019}. PSP observations in particular show patches of extremely large amplitude waves ($\delta B/B \sim 1$) that can even reverse the direction of the (mainly radial) background magnetic field: for this reason, these structures have been dubbed "switchbacks", and their morphology, formation, and evolution are current topics of active observational\citep{bale2019,kasper2019,horbury2020,laker2020,krasnoselskikh2020,larosa2020,bale2021,tenerani2021} and theoretical\citep{squire2020,zank2020,drake2020,tenerani2020,drake2021,mallet2021,schwadron2021,johnston2022,squire2022a,squire2022b} interest.

The inspiration for the current paper is the fact that, while the large-amplitude MHD AW appears to describe many observed aspects of the switchbacks, these structures also often have remarkably sharp, discontinuous boundaries\citep{farrell2020,farrell2021}, with large-amplitude rotations of the magnetic field over only a few ion inertial lengths $d_i=v_A/\Omega_i$, where $\Omega_i = Ze B/m_ic$ is the ion gyrofrequency. The formation of these sharp boundaries is an open topic: it may be a required by-product of enforcing the constant magnetic field strength condition \citep{squire2022b}. Nevertheless, the MHD model cannot reasonably be expected to apply on these scales, and this provides motivation for a theoretical treatment of large-amplitude "kinetic Alfv\'en waves". Moreover, there appear to be localised decreases in the magnetic field strength and associated density changes at the switchback boundaries\cite{farrell2020}, which suggest a modification of the Alfv\'en waves on these small scales.

Linear properties of the KAW are well-known\cite{hasegawa1976b,lysak1996,stasiewicz2000,hollweg1999}. Including dispersion due to finite frequency ($\omega/\Omega_i \sim \kpar d_i$) and at the ion sound radius $\rho_s = \sqrt{ZT_e/m_i}/\Omega_i$. Assuming very oblique waves, $\kperp \gg \kpar$, the linear dispersion relation is
\beq
\omega = \pm \kpar \vA \sqrt{\frac{1+\kperp^2\rho_s^2}{1+\kpar^2 d_i^2}}.\label{eq:disp}
\eeq
The AW also linearly develops a density fluctuation, given by
\beq
\frac{\delta n}{\avg{n}} = - i \kperp d_i \frac{\kpar \vA}{\omega}\frac{\delta B}{\avg{B}}.\label{eq:dens}
\eeq
An important point made by Hollweg\cite{hollweg1999} is that this density fluctuation becomes significant at $\kperp d_i\sim 1$: for $\kpar/\kperp \ll1$ and $\beta =\rho_s^2/d_i^2 \ll 1$ (both typical of the corona), this occurs at a much larger scale than that at which the corrections to the dispersion relation (\ref{eq:disp}) become relevant. 

Because the Alfv\'en velocity (\ref{eq:vadef}) depends on the density, one might expect these density fluctuations (and/or associated compressive flows) to drive nonlinear steepening, since the wave's velocity now depends on the phase. Moreover, there may exist steady nonlinear waves, in which nonlinear steepening is balanced by the dispersion (\ref{eq:disp}): if these disturbances have a finite spatial extent, we will (loosely) call them \emph{solitons}\footnote{Loosely, since we do not here study the interactions between them to prove that they survive unscathed, part of the usual definition.}. The calculation we present here is a detailed examination of this process and the resulting solitons.

Much work on nonlinear kinetic Alfv\'en waves has been carried out in the "gyrokinetic" small-amplitude, low-frequency limit\cite{schektome2009,zocco2011}: in this regime, steepening is ordered out. These models have been spectacularly successful in predicting many properties of the small-scale, small-amplitude kinetic turbulence observed in the solar wind \citep{chen2013,duan2021} and in numerical simulations \citep{grovselj2018}; on the other hand, large-amplitude structures like switchbacks and their sharp edges are firmly outside their regime of validity.

Going beyond this, there are two main bodies of work on nonlinear steepening of oblique one-dimensional Alfv\'en waves and associated solitons \markup{(there is also a large literature on small amplitude parallel or quasi-parallel Alfv\'en waves\cite{cohen1974,rogister1971,mjolhus1986}; since this is not applicable to our focus here, we will not discuss this further).} The first group (not chronologically) began with the analysis of Hasegawa \& Mima\cite{hasegawa1976}, whose calculation predicted the existence of arbitrary-amplitude exact kinetic Alfv\'en solitons with density spikes, with parallel extent much longer than $d_i$ and perpendicular extent of order $\rho_s$; these solitons involved magnetic fluctuations only in the direction perpendicular to both the background magnetic field and the direction of propagation. A similar calculation was carried out by Shukla \emph{et al.}\cite{shukla1982}, but for inertial Alfv\'en solitons\markup{, for nonlinear kinetic Alfv\'en waves in a two-electron-temperature plasma by Berthomier et al. 1999\cite{berthomier1999}, with the inertial and kinetic Alfv\'en solitons unified by Wu et al. 1996\cite{wu1996}. A later extension of these theories was also made, attempting to include the dissipative effects of turbulence by means of an effective collisionality by Wu 2003\cite{wu2003}, while the whole theory has been reviewed concisely by Wu \& Chao 2004\cite{wu2004}. However, concurrently} Seyler \& Lysak\cite{seyler1999} pointed out that these calculations neglected a nonlinear term in their derivation \markup{(we have confirmed that the subsequent publications we listed above also neglect this term)}: including this nonlinear term, the only steady nonlinear solitons necessarily contained discontinuous density profiles, calling their existence into question. It should be noted that all of these works did not provide a fully systematic derivation of their equations, and implicitly used a small-amplitude approximation. In a previous paper, we have systematically (re)-derived\cite{mallet2023} the equations used by Seyler \& Lysak, providing support for their conclusion that small-amplitude kinetic and/or inertial Alfv\'en solitons with continuous density profiles do not exist.

Quite different to the above, the second approach (in fact, chronologically the older) is that of Kakutani \& Ono\cite{kakutani1969}. Attempting to find a weakly nonlinear dispersive equation governing Alfv\'en waves in a cold plasma, with arbitrary propagation angle $\theta$ relative to the mean field, they found that for small amplitude waves, the standard procedure failed, and just produced a linear dispersive equation. This is because the AW can avoid steepening by, as mentioned above, maintaining a constant magnetic field strength (Eq.~\ref{eq:stconst}). To obtain a nonlinear equation, they were forced to take a \emph{large}-amplitude AW at lowest order, eventually deriving the modified Korteweg-de Vries (mKdV) equation,
\beq
\frac{1}{\cos\theta}\partial_{T} \tn + \frac{1}{2}\cot^2\theta \ddxi^3 \tn + \frac{3}{4}\cot^2\theta \tn^2\ddxi \tn,
\eeq
where we have ignored electron inertia for simplicity, $\tn=\delta n/n_0$, with $T=\Omega_i t$ and  $\xi = x/d_i$. This equation is quite peculiar, in that both the nonlinear and linear terms depend on $\cot^2\theta$, which arises from (weak) dispersion due to finite $\kpar d_i$. This results in both density spike and dip solitons which necessarily involve a full rotation of the magnetic field vector in the plane transverse to the direction of propagation: they are necessarily large-amplitude,  not at all linearly polarized, and have widths much longer than $d_i$, fundamentally different from the approach of Hasegawa \& Mima\cite{hasegawa1976}, Shukla \emph{et al.}\cite{shukla1982}, and Seyler \& Lysak\cite{seyler1999} mentioned above. This work was then extended to include the effect of isothermal electrons by Kawahara\cite{kawahara1969}, who also found the mKdV equation with different coefficients (see Eq.~\ref{eq:mkdv} in this paper). \markup{The exact solitary solutions of the two-fluid system have also been investigated in detail by Dubinin et al. 2005\citep{dubinin2005} and Mj\"olhus 2006\cite{mjolhus2006}, confirming the existence of oblique Alfv\'en solitons: however, these works are unable to provide a simple, intuitive, dynamical equation like the mKdV equation.}

In this work, we will unify these two approaches, deriving a weakly nonlinear evolution equation for slightly dispersive, very oblique AW at small $\beta$, applicable for small scale AW with lengthscales comparable to $d_i$. We will be able to derive generalizations of the mKdV solitons\cite{kakutani1969} to the case where the density fluctuations become large, $\delta n/n_0\sim 1$ at the $d_i$ scale, and also recover the discontinuous, small-amplitude solutions due to Seyler \& Lysak\cite{seyler1999}. We will discuss the relevance of our equation and its solutions for switchback edges and for imbalanced turbulence in the solar corona\markup{ as well as its limitations: since for simplicity we use a fluid model with isothermal electrons and cold ions, we expect our solutions to be modified by finite ion temperature effects, as well as kinetic effects such as a realistic FLR response and wave-particle interactions such as Landau and cyclotron damping. Extensions of this work to take into account these important processes may be able to better match the observed properties of the switchbacks.}
\section{Basic equations}
Our starting point are one-dimensional, quasineutral two-fluid equations with isothermal electrons and cold ions, neglecting electron inertia\footnote{Mathematically, the inertial dispersion can be included without too much trouble, but since when $\rho_s\sim d_e$ kinetic effects like Landau damping become important we will neglect it.} and collisions.  Without loss of generality, we assume propagation in the $x$ direction, and that the mean magnetic field is in the $x-z$ plane, i.e. 
\beq
\boldsymbol{v}_{\rm A}= (\vA\cos\theta,0,\vA\sin\theta).\label{eq:vacomp}
\eeq
Eliminating the electric field, electron velocity and electron density in favour of the other variables, these may be written
\begin{align}
    \ddt n &= -\ddx(n u_x),\label{eq:dtnmain}\\
    \Dt u_x &= - \frac{1}{n}\ddx \left(c_s^2 n + \frac{1}{2}b_y^2 + \frac{1}{2}b_z^2\right),\label{eq:dtuxmain}\\
    \Dt u_y &=\frac{1}{n}\vA\cos\theta\ddx b_y,\label{eq:dtuymain}\\
    \Dt u_z &=\frac{1}{n}\vA\cos\theta\ddx b_z,\label{eq:dtuzmain}\\
    \ddt b_y &=\vA\cos\theta \left[\ddx u_y + d_i\ddx\left(\frac{1}{n}\ddx b_z\right)\right]-\ddx[u_x b_y],\label{eq:dtbymain}\\
    \ddt b_z &=\vA\cos\theta \left[\ddx u_z - d_i\ddx\left(\frac{1}{n}\ddx b_y\right)\right]-\ddx[u_x b_z],\label{eq:dtbzmain}
\end{align}
where $n=n_i/\avg{n}$ is the ion density normalized to its mean value, $\vu$ is the ion velocity, $\vb=\vB/\sqrt{4\pi \avg{n}m_i}$ is the magnetic field in Alfv\'en units (note that $b_x = \vA\cos\theta$ is a constant since the propagation is one-dimensional), $c_s = \sqrt{ZT_e/m_i}$ is the sound speed, and $\Dt = \ddt + u_x \ddx$.

Assuming that the amplitude of the fluctuations is small enough that all nonlinear terms may be neglected, and supposing all fluctuations vary proportional to $\exp(i(kx-\omega t)$, we obtain the linear dispersion relation
\begin{align}
(\kpar^2\vA^2 - \omega^2)&\left[(\kpar^2\vA^2-\omega^2)(\omega^2-k^2c_s^2)+\omega^2 k_\perp^2\vA^2\right] \nonumber\\&= (\omega^2-k^2c_s^2)\frac{\omega^2}{\Omega_i^2}k^2v_A^2 \kpar^2\vA^2,\label{eq:dispfull}
\end{align}
where $\kpar=k\cos\theta$ and $k_\perp=k\sin\theta$. Taking $\Omega_i\to\infty$ while keeping $\omega\sim k\vA\sim k c_s$ constant, we may neglect the RHS of (\ref{eq:dispfull}), which obviously leaves the MHD dispersion relation with non-dispersive Alfv\'en, slow, and fast modes. Setting $c_s^2=0$, we obtain a mode with $\omega=0$ and the remaining dispersion relation
\beq
(k^2\vA^2-\omega^2)(\kpar^2\vA^2-\omega^2) - \frac{\omega^2}{\Omega_i^2}k^2\vA^2\kpar^2\vA^2=0,
\eeq
which encodes the Alfv\'en-ion cyclotron and whistler waves.

Taking 
\beq
\omega\sim \kpar\vA, k\rho_s \sim 1\quad \kpar/k\sim \sqrt{\beta}\ll1,\label{eq:kaworderhighbeta}
\eeq
which together imply $k d_i \gg1$ and $\kpar d_i \sim 1$, we order out the whistlers and recover the previously mentioned KAW dispersion relation (\ref{eq:disp})\markup{: since in this limit $\kpar\ll k$, we may approximate $k\approx k_\perp$.}

In the calculation presented in this paper we will be interested in AW that are only slightly dispersive. We take
\beq
\omega^2 = \kpar^2\vA^2 + \Delta,
\eeq
with 
\beq
\Delta\sim \kpar^2 d_i^2 \sim k^2\rho_s^2\ll1,\label{eq:slightkaworderlin}
\eeq
i.e. an Alfv\'en wave that is only slightly dispersive. We also take $\sin^2\theta\gg\Delta$, i.e. the wave is not extremely parallel. Then, (\ref{eq:dispfull}) to first order in $\Delta$ gives
\beq
\Delta = \frac{1}{\sin^2\theta}\left(k^2\rho_s^2-\kpar^2d_i^2\right)\kpar^2\vA^2, \label{eq:delta}
\eeq
so that in this limit the Alfv\'en wave frequency is
\beq
\omega = \pm \kpar\vA\left(1+\frac{k^2\rho_s^2-\kpar^2d_i^2}{2\sin^2\theta}\right)\label{eq:slightkawdisp}
\eeq
Evidently, the oblique limit ($\sin^2\theta-1\ll1$) of this agrees with the slightly dispersive limit of (\ref{eq:disp}).

These linear properties are quite well-known: in the rest of the paper, we focus on the nonlinear properties of the slightly dispersive AW.

\section{Two timescale expansion}
To study large-amplitude, nonlinear AW with transverse scales of order $d_i$, we take
\beq
\frac{b_y}{\vA}\sim \frac{b_z}{\vA}\sim d_i\ddx \sim n-1\sim 1, \quad \cos^2\theta \sim \beta \sim \epsilon \ll1.\label{eq:ordering}
\eeq
Note that the density fluctuation $n-1$ is also large: linearly, the density fluctuation is proportional to $k d_i$ (see Eq.~\ref{eq:dens}),\cite{hollweg1974} so if we want large-amplitude magnetic fluctuations with gradients on scales of order $d_i$, we must allow this. Moreover, this ordering means that 
\beq
\rho_s^2 \sim \dpar^2 \sim \epsilon d_i^2,
\eeq
where we define the shorthand $\dpar^2=d_i^2\cos^2\theta$, and that
\beq
c_s^2 \sim \vA^2\cos^2\theta \sim \epsilon \vA^2.
\eeq
This means that our solution will be only slightly dispersive. For direct applicability of our results to the solar wind and corona, the most dubious of the assumptions going into our model is that $\beta$ is small (moreover, we have assumed that the ions are very cold so that we can ignore FLR effects): the solar wind typically has $\beta\sim1$. Thus, our results may only apply directly somewhat lower down in the corona where $\beta\sim 0.01$, and may be significantly modified by the presence of wave-particle interactions like Landau damping\citep{medvedev1996}. \markup{Additionally, our neglect of electron inertia means that $\beta_e$ cannot become smaller than the mass ratio; the (fairly wide) range of $\beta_e$ in which our analysis is valid is $1 \gg \beta_e\gg m_e/m_i$.}

We expand the variables and equations in powers of $\epsilon$,
\begin{align}
    b_y = b_{0y}+ &\epsilon b_{1y}+\ldots,\nonumber\\
    b_z = b_{0z}+ &\epsilon b_{1z}+\ldots,\nonumber\\  
    u_y = u_{0y}+ &\epsilon u_{1y}+\ldots,\nonumber\\
    u_z = u_{0z}+ &\epsilon u_{1z}+\ldots,\nonumber\\
    u_x = u_{0x}+&\epsilon u_{1x}+\ldots,\nonumber\\
    n = n_0+ &\epsilon n_1+\ldots,
\end{align}
where $b_{z}$ includes the transverse background field $\vA\sin\theta$. 
At first order, the naive expansion will fail and we must introduce a slow timescale $\tau=\epsilon t$, so that our variables depend on
\beq
f=f(x,t,\tau),
\eeq
and
\beq
\left.\frac{\dd }{\dd t}\right|_x = \left.\frac{\dd }{\dd t}\right|_{x,\tau} + \epsilon\left.\frac{\dd}{\dd \tau}\right|_{x,t}.
\eeq
For brevity we will use the notation
\beq
\ddt = \left.\frac{\dd }{\dd t}\right|_{x,\tau}, \quad \ddtau = \epsilon\left.\frac{\dd}{\dd \tau}\right|_{x,t}.
\eeq
We also want the fast time evolution to have a particular form:  the lowest order solution should be an Alfv\'en wave, so the dependence on $x$ and $t$ is required to be
\beq
f=f(x-t\vA\cos\theta,\tau).
\eeq
Then, 
\beq
\ddt = -\vA\cos\theta\ddx.
\eeq
We will apply one of two types of boundary conditions. First, for localized fluctuations,
\begin{align}
    &b_{0y}\to b_{y\pm},\quad
    b_{0z}\to b_{z\pm},\nonumber\\
   & u_{0y}\to u_{y\pm},\quad
    u_{0z}\to u_{z\pm},\nonumber\\
    &u_{0x}\to 0,\quad
    n_0\to 1, \nonumber \\ \quad &\text{as } x\to\pm\infty,
\end{align}
while fluctuations at first order and higher are all required to vanish as $|x|\to\infty$. By allowing different constants at $\pm\infty$, we are in principle allowing finite kinks in the magnetic field. The constants do however need to be related to each other (see later; Eq.~\ref{eq:constblocalized}).

Alternatively, we can use periodic boundary conditions. We define the spatial average
\beq
\avg{f}(t,\tau)=\frac{1}{\Lambda}\int_0^\Lambda f(x,t,\tau) dx,\label{eq:avgdef}
\eeq
where $\Lambda$ is the periodicity length. We allow $\avg{b}_{0y},\avg{b}_{0z}$ to be constants, with $\avg{b}_{0y}^2+\avg{b}_{0z}^2=\vA^2\sin^2\theta$, while $\avg{n}_0=1$, and all other averages of single variables vanish.

We can now proceed with our expansion of the equations.

\subsubsection{$O(1)$}
From (\ref{eq:dtnmain}), we obtain 
\beq
-\vA\cos\theta\ddx n_0 = -\ddx(n_0 u_{0x}).
\eeq
Integrating once,
\beq
u_{0x} = \vA\cos\theta \frac{n_0-1}{n_0},\label{eq:u0x}
\eeq
where we have used assumed either localised or periodic fluctuations to set the constant of integration. This is somewhat small; assuming $n_0\sim n_0-1\sim 1$, we have $u_{0x} \sim \epsilon^{1/2}\vA$. Using this in (\ref{eq:dtuxmain}), we find that the $\Dt u_{x}$ and $(1/n)c_s^2\ddx n$ terms are in fact $O(\epsilon)$ by virtue of our ordering (\ref{eq:ordering}). After integrating the remaining term, we have
\beq
b_{0y}^2 + b_{0z}^2 = b_T^2,\label{eq:constb}
\eeq
where $b_T^2$ is a constant, the transverse magnetic field strength. In the case of localised fluctuations, 
\beq
b_T^2 = \vA^2\sin^2\theta = b_{y+}^2 + b_{z+}^2 = b_{y-}^2 + b_{z-}^2, \label{eq:constblocalized}
\eeq
but for periodic fluctuations $b_T^2$ need not be equal to $\vA^2\sin^2\theta$\footnote{We could force this to be the case; it amounts to redefining $\vA$ and $\theta$ while keeping the phase speed $\vA\cos\theta$ constant, but it has no dynamical significance.}. Turning now to (\ref{eq:dtuymain}), upon inserting (\ref{eq:u0x}) we find
\beq
-\vA\cos\theta\left(\frac{n_0-(n_0-1)}{n_0}\right)\ddx u_{0y} = \frac{1}{n_0}\vA\cos\theta\ddx b_{0y}.
\eeq
The nonlinearity due to $u_{0x}$ on the LHS and due to $1/n_0$ on the RHS of (\ref{eq:dtuymain}) therefore cancels out. Since $n_0>0$ we can integrate once, obtaining
\beq
u_{0y}=-b_{0y},\label{eq:u0y}
\eeq
where we have chosen the integration constant to be zero without loss of generality for both sets of boundary conditions. Similarly, from (\ref{eq:dtuzmain}) we find
\beq
u_{0z}=-b_{0z}.
\eeq
Using (\ref{eq:u0y}) in (\ref{eq:dtbymain}), two terms cancel and we are left with
\beq
d_i\vA\cos\theta\ddx\left(\frac{1}{n_0}\ddx b_{0z}\right)=\vA\cos\theta\ddx\left(\frac{n_0-1}{n_0}b_{0y}\right).
\eeq
Integrating once,
\beq
(n_0-1)b_{0y} = d_i\ddx b_{0z}+C_y.\label{eq:n0b0y}
\eeq
Similarly, from (\ref{eq:dtbzmain}), we obtain after cancellation and integrating once
\beq
(n_0-1)b_{0z} = -d_i\ddx b_{0y}+C_z.\label{eq:n0b0z}
\eeq

To sum up our results so far, we have found that in this regime, the lowest-order solution is a constant-$B^2$ AW, just as in MHD. However, since $d_i\ddx \sim 1$, this AW also has large density fluctuations. Due to the constancy of the magnetic field strength, it is worth introducing the angle $\Phi$ that the zeroth-order transverse magnetic field makes in the $y$-$z$ plane,
\beq
b_{0y}=b_T\sin\Phi,\quad b_{0z}=b_T\cos\Phi.
\eeq
Usefully, we have
\begin{align}
\ddx b_{0y} &= b_{0z}\ddx\Phi, \quad \ddx b_{0z}=-b_{0y}\ddx\Phi.\nonumber\\
\ddtau b_{0y} &= b_{0z}\ddtau\Phi, \quad \ddtau b_{0z}=-b_{0y}\ddtau\Phi.\label{eq:phirels}
\end{align}
We may use these in (\ref{eq:n0b0y}) and (\ref{eq:n0b0z}) to find
\begin{align}
n_0-1 &= -d_i\ddx\Phi +\frac{C_y}{b_T\sin\Phi},\\
n_0-1 &= -d_i\ddx\Phi +\frac{C_z}{b_T\cos\Phi},
\end{align}
whence $C_y=C_z=0$, so that
\beq
n_0 =1-d_i\ddx\Phi.\label{eq:n0phi}
\eeq
Since $\infty>n_0>0$, for our solution to be valid $-\infty<d_i\ddx\Phi<1$. Indeed, if $n_0$ were to become as small as $\epsilon$, terms involving $n_1$ would be relevant in the lowest order equations, and our expansion will fail.
\subsubsection{$O(\epsilon)$}
At this order, the (nonlinear) dispersive terms enter and the wave slowly evolves in time. Eq.~(\ref{eq:dtnmain}) gives
\beq
\ddtau n_0 - \vA\cos\theta \ddx n_1 = -\ddx(n_1 u_{0x}+n_0u_{1x}).
\eeq
Using (\ref{eq:n0phi}), we may integrate this once; also substituting for $u_{0x}$ using (\ref{eq:u0x}),
\beq
u_{1x} = \vA\cos\theta \frac{n_1}{n_0^2} + d_i\frac{\ddtau\Phi}{n_0} + \frac{C_{n}}{n_0},\label{eq:u1x}
\eeq
where $C_n$ is the constant of the integration. From (\ref{eq:dtuxmain}), we obtain
\begin{align}
-\vA\cos\theta\ddx u_{0x}&+u_{0x}\ddx u_{0x} +\frac{1}{n_0}c_s^2\ddx n_0\nonumber\\& = -\frac{1}{n_0}\ddx(b_{0y}b_{1y} + b_{0z}b_{1z}).
\end{align}
Inserting (\ref{eq:u0x}) and multiplying by $n_0>0$,
\beq
\vA^2\cos^2\theta \ddx\left(\frac{1}{n_0}\right) + c_s^2\ddx n_0 = -\ddx(b_{0y}b_{1y} + b_{0z}b_{1z}),
\eeq
which relates the density fluctuation to the first-order magnetic-field-strength fluctuations. This may also be written
\beq
\left(\beta-\cos^2\theta\frac{1}{n_0^2}\right)\ddx n_0 = -\frac{1}{\vA^2}\ddx(b_{0y}b_{1y} + b_{0z}b_{1z}).\label{eq:n0b0b1}
\eeq
At first order, (\ref{eq:dtuymain}) gives
\begin{align}
\ddtau u_{0y}+&(u_{0x}-\vA\cos\theta)\ddx u_{1y} + u_{1x}\ddx u_{0y} \nonumber\\&= \frac{1}{n_0}\vA\cos\theta\ddx b_{1y}- \frac{n_1}{n_0^2}\vA\cos\theta\ddx b_{0y}.
\end{align}
Using (\ref{eq:u0y}), (\ref{eq:u0x}) and (\ref{eq:u1x}), we write this as
\beq
-\ddtau b_{0y} - \frac{1}{n_0}\vA\cos\theta\ddx(u_{1y}+b_{1y}) = \left[\frac{d_i}{n_0}\ddtau\Phi+\frac{C_n}{n_0}\right]\ddx b_{0y},
\eeq
and now using (\ref{eq:phirels}), we obtain
\beq
\ddtau b_{0y} = -\vA\cos\theta\ddx(u_{1y}+b_{1y})- C_n\ddx b_{0y}.\label{eq:dtaub0y}
\eeq
Similar manipulations of the first order part of (\ref{eq:dtuzmain}) give
\beq
\ddtau b_{0z}=-\vA\cos\theta\ddx(u_{1z}+b_{1z})-C_n\ddx b_{0z}.
\eeq
The first order piece of (\ref{eq:dtbymain}) is
\begin{align}
\ddtau b_{0y}&-\vA\cos\theta\ddx b_{1y} \nonumber\\&= \vA\cos\theta\ddx u_{1y}+ d_i\vA\cos\theta\ddx\left(\frac{1}{n_0}\ddx b_{1z} - \frac{n_1}{n_0^2}\ddx b_{0z}\right) \nonumber\\&\quad- \ddx\left(u_{0x}b_{1y}+ u_{1x}b_{0y}\right),
\end{align}
where we have first used (\ref{eq:dtuzmain}) to replace $u_z$ in favor of $b_z$. Adding this equation to (\ref{eq:dtaub0y}) to eliminate $u_{1y}$ and $b_{1y}$, and using (\ref{eq:u0x}) and (\ref{eq:u1x}),
\begin{align}
2\ddtau b_{0y} = &-C_n\ddx \left(\frac{1+n_0}{n_0}b_{0y}\right) \nonumber\\&+ d_i\vA\cos\theta\ddx\left(\frac{1}{n_0}\ddx b_{1z} - \frac{n_1}{n_0^2}\ddx b_{0z}\right)\nonumber\\&-\vA\cos\theta\ddx\left(\frac{n_0-1}{n_0}b_{1y} + \frac{n_1}{n_0^2}b_{0y}\right)\nonumber\\&-d_i\ddx\left(\frac{b_{0y}}{n_0}\ddtau\Phi\right).\label{eq:dtaub0y2}
\end{align}
Now, we define
\beq
h_{0y}=\int_{x_0}^x b_{0y}dx + H_y,
\eeq
where $H_y$ is a constant and $x_0$ is an arbitrary constant position, and integrate (\ref{eq:dtaub0y2}) with respect to $x$, obtaining
\begin{align}
2\ddtau h_{0y} = &-C_n\frac{1+n_0}{n_0}b_{0y}\nonumber\\&+ d_i\vA\cos\theta \left(\frac{1}{n_0}\ddx b_{1z} - \frac{n_1}{n_0^2}\ddx b_{0z}\right)\nonumber\\
&-\vA\cos\theta\left(\frac{n_0-1}{n_0}b_{1y}+\frac{n_1}{n_0^2}b_{0y}\right)\nonumber\\
&+\frac{d_i}{n_0}\ddtau b_{0z},\label{eq:dtaub0y3}
\end{align}
where we can absorb the constant of integration into $H_y$. Similar manipulations on the first-order piece of (\ref{eq:dtbzmain}) give
\begin{align}
2\ddtau h_{0z}=&-C_n\frac{1+n_0}{n_0}b_{0z}\nonumber\\&-d_i\vA\cos\theta \left(\frac{1}{n_0}\ddx b_{1y} - \frac{n_1}{n_0^2}\ddx b_{0y}\right)\nonumber\\
&-\vA\cos\theta\left(\frac{n_0-1}{n_0}b_{1z}+\frac{n_1}{n_0^2}b_{0z}\right)\nonumber\\
&-\frac{d_i}{n_0}\ddtau b_{0y},\label{eq:dtaub0z2}
\end{align}
where
\beq
h_{0z}=\int_{x_0}^x b_{0z}dx + H_z.
\eeq
Now, we form the following combination of (\ref{eq:dtaub0y3}) and (\ref{eq:dtaub0z2}):
\begin{widetext}
\begin{align}
    &2b_{0y}\ddtau h_{0z}-2b_{0z}\ddtau h_{0y}\nonumber\\&= - C_n \frac{n_0+1}{n_0}\cancel{\left[b_{0y}b_{0z}-b_{0z}b_{0y}\right]}\nonumber\\
    &-d_i\vA\cos\theta\left[\frac{1}{n_0}\left(b_{0y}\ddx b_{1y}+b_{0z}\ddx b_{1z}\right)-\frac{n_1}{n_0^2}\cancel{\left(b_{0y}\ddx b_{0y} + b_{0z}\ddx b_{0z}\right)}\right]\nonumber\\
    &-\vA\cos\theta\left[\frac{n_0-1}{n_0}\left(b_{1z}b_{0y}-b_{1y}b_{0z}\right)+\frac{n_1}{n_0^2}\cancel{\left(b_{0y}b_{0z}-b_{0z}b_{0y}\right)}\right]\nonumber\\
    &-\frac{d_i}{n_0}\cancel{\left[b_{0y}\ddtau b_{0y}+b_{0z}\ddtau b_{0z}\right]},
\end{align}
\end{widetext}
where for clarity we have written out various terms which cancel. Recalling (\ref{eq:n0phi}) and (\ref{eq:phirels}), the remaining terms combine into
\beq
b_{0y}\ddtau h_{0z}-b_{0z}\ddtau h_{0y} = -\frac{1}{2}\frac{d_i\vA\cos\theta}{n_0}\ddx(b_{0y}b_{1y}+b_{0z}b_{1z}).
\eeq
The RHS involves the gradient of the first-order magnetic field strength, which is related to the density by (\ref{eq:n0b0b1}), so that
\begin{align}
b_{0y}\ddtau h_{0z}-b_{0z}\ddtau h_{0y} &= \frac{1}{2}d_i\vA^3\cos\theta\left(\frac{\beta n_0^2 - \cos^2\theta}{n_0^3}\right)\ddx n_0\nonumber\\&=-\mu\label{eq:mustep0}
\end{align}
defining the function $\mu$. Differentiating (\ref{eq:mustep0}) with respect to $x$, we obtain
\beq
\left(b_{0z}\ddtau h_{0z} + b_{0y}\ddtau h_{0y}\right)\ddx\Phi= F,
\eeq
where
\beq
F=b_T^2\ddtau \Phi - \ddx \mu.\label{eq:Fdef}
\eeq
Differentiating with respect to $x$ again,
\beq
\frac{F\ddx^2\Phi}{\ddx\Phi} + (\ddx \Phi)^2\mu = \ddx F,
\eeq
where we have cancelled some terms due to the constancy of $b_T^2$. Integrating this equation once over $x$,
\beq
F=\ddx\Phi\int \mu\ddx\Phi dx + C_\Phi\ddx\Phi,\label{eq:Feq}
\eeq
where $C_\Phi$ is a constant. Let us temporarily retreat to the small-amplitude wave in order to determine it. The first term on the RHS is a purely nonlinear term, and in the small-amplitude limit it may be neglected. Moreover, at small amplitude, $b_T^2\approx \vA^2\sin^2\theta$. Inserting $\Phi\propto \exp(i(kx-\delta\tau))$, and expanding $\mu$ in $n_0-1\ll1$, we find using the definition of $F$ that in this limit
\beq
\delta = \frac{\kpar\vA(k^2\rho_s^2-k^2\dpar^2)}{2\sin^2\theta} -\frac{kC_\Phi}{\vA^2\sin^2\theta},
\eeq
which agrees with the dispersion relation (\ref{eq:slightkawdisp}), provided that we set $C_\Phi=0$. It will be more convenient from now on to work in terms of $n_0$ (see \ref{eq:n0phi}), so we differentiate (\ref{eq:Feq}) with respect to $x$, also using (\ref{eq:Fdef}), to obtain
\beq
\ddtau n_0 =\ddx^2\hmu + \frac{1}{d_i^2}\ddx\left[(n_0-1)\int (n_0-1)\hmu dx\right],
\eeq
where
\beq
\hmu = -\frac{d_i \mu}{b_T^2} = \frac{1}{2}\vA\cos\theta\frac{\vA^2}{b_T^2}\left(\frac{\rho_s^2 n_0^2 - \dpar^2}{n_0^3}\right)\ddx n_0.
\eeq
Now we will write the whole equation in terms of $n_0$. First,
\beq
\ddx^2\hmu = \frac{1}{2}\vA\cos\theta\frac{\vA^2}{b_T^2}\ddx^3\left(\rho_s^2\log n_0 + \dpar^2\frac{1}{2n_0^2}\right).
\eeq
This term is both dispersive and nonlinear. To calculate the term involving the integral, note that
\begin{align}
\frac{1}{d_i^2}&(n_0-1)\hmu \nonumber\\&= \frac{1}{2}\vA\cos\theta \frac{\vA^2}{b_T^2}\ddx\left[\beta(n_0-\log n_0)+\cos^2\theta\frac{2n_0-1}{2n_0^2}\right].
\end{align}
Integrating this, we find
\begin{align}
&\frac{1}{d_i^2}(n_0-1)\int (n_0-1)\hmu dx\nonumber\\&=\frac{1}{2}\vA\cos\theta\frac{\vA^2}{b_T^2}(n_0-1)\left[\beta(n_0-\log n_0)+\cos^2\theta\frac{2n_0-1}{2n_0^2}+C_I\right].
\end{align}
where $C_I$ is the constant of integration. To ensure compatibility with the dispersion relation (the same argument that led to $C_\Phi=0$) the coefficient of $n_0-1$ in the expansion for $n_0-1\ll1$ must vanish, so that
\beq
C_I=-\beta-\frac{\cos^2\theta}{2}.
\eeq
Then,
\begin{align}
&\frac{1}{d_i^2}(n_0-1)\int (n_0-1)\hmu dx\nonumber\\&=\frac{1}{2}\vA\cos\theta\frac{\vA^2}{b_T^2}\left[\beta\left\{(n_0-1)^2-(n_0-1)\log n_0\right\}\phantom{\frac{(n_0-1)^3}{2n_0^2}}\right.\nonumber\\&\left.\quad\quad\quad\quad\quad\quad\quad-\cos^2\theta\frac{(n_0-1)^3}{2n_0^2}\right].
\end{align}
Finally, we define normalised variables
\beq
\zeta= \frac{\beta\vA^3\cos\theta}{2 d_i b_T^2}t, \quad \xi = \frac{x}{d_i}, \quad \gamma=\frac{\cos^2\theta}{\beta}
\eeq
and our nonlinear equation is
\begin{widetext}
\beq
\ddze n_0 = \ddxi^3\left[\log n_0 + \frac{\gamma}{2n_0^2}\right]+ \ddxi\left[(n_0-1)^2-(n_0-1)\log n_0 - \gamma\left\{\frac{(n_0-1)^3}{2n_0^2}\right\}\right].\label{eq:master}
\eeq
\end{widetext}
In terms of $\xi$,
\beq
n_0=1-\ddxi \Phi.
\eeq
\markup{It is also worth pointing out that $\gamma = \dpar^2/\rho_s^2$: physically, this parameter controls the relative importance of the two dispersive terms in the linearized AW dispersion relation, Eq.~\ref{eq:disp}.}
\subsection{$d_i\ddx\ll1$; the mKdV equation}
Keeping the magnetic fluctuations (i.e. $\Phi$) at the same large amplitude while taking $d_i\ddx=\ddxi\ll1$, $n_0-1$ becomes small but the dispersive and purely nonlinear terms are comparable; in fact, at lowest order in $\tilde{n} = n_0-1\ll1$, we recover the modified Korteweg-de Vries (mKdV) equation \citep{kakutani1969,kawahara1969},
\beq
\ddze \tilde{n} = (1-\gamma)\ddxi^3\tilde{n} + \frac{1}{2}(1-\gamma)\ddxi(\tilde{n}^3),\label{eq:mkdv}
\eeq
as expected.\markup{To be precise, the calculation of Kakutani \& Ono\cite{kakutani1969} was for cold plasma, and both that work and  Kawahara\cite{kawahara1969} included electron inertia, so the coefficients in their mKdV equations are slightly different: exact agreement is obtained by taking $m_e/m_i=0$ in the equation of Kawahara\cite{kawahara1969}. We should also point out that the mKdV equation does not require $\beta\sim\cos^2\theta\ll1$; just $d_i\ddx\ll1$.} In this limit ($\ddxi\ll1$), the nonlinear and dispersive terms are comparable at a scale $L$ if
\beq
\tilde{n}\sim \frac{d_i}{L}.
\eeq
In terms of the fluctuation in $\Phi$ across the scale $L$, $\delta\Phi$,
\beq
\delta\Phi \sim 1,
\eeq
independent of $L$. This is quite a peculiar situation: the nonlinearity only becomes relevant if the waves have extremely large amplitude; for smaller-amplitude waves, the nonlinearity is irrelevant and waves will (slowly) disperse. Moreover, it is worth pointing out that $\delta\Phi\gg1$ does \emph{not} correspond to $\delta B/B\gg1$: in fact, it corresponds to circularly polarized AW with increasingly short wavelength as $\delta \Phi$ increases for fixed $L$. Thus, we conclude that for a wave with magnetic-field fluctuations on a scale $L$, the nonlinear steepening is only marginally important when $\delta b/b_T\sim1$. This is borne out by (for example) the mKdV solitons,
\begin{align}
    n &= 2a\mathrm{sech} a(\xi-\lambda\zeta)\nonumber\\
    \Phi&=4\tan^{-1}\mathrm{tanh}\frac{a(\xi-\lambda\zeta)}{2},
\end{align}
where $a=\pm\sqrt{-\lambda/(1-\gamma)}$, $\lambda$ being the soliton velocity in the $\zeta$-$\xi$ units, which is required to have the opposite sign to $1-\gamma$ for a solitary solution. No matter the width of the soliton ($1/a$), the transverse magnetic field undergoes a full rotation by $2\pi$ across it. In other words, according to the mKdV equation, solitary structures whose transverse magnetic field does not undergo a complete rotation should disperse over time.
\subsection{Magnetic field strength fluctuations}
Observationally, it is often the case that the magnetic field is more readily observable on small scales than the density. To find the magnetic field strength fluctuations, we integrate (\ref{eq:n0b0b1}), yielding
\beq
\frac{\delta |b|^2}{\vA^2} = -\beta\left(n_0+\frac{\gamma}{n_0}\right)+\beta(1+\gamma),\label{eq:magfieldstrength}
\eeq
where we have set the constant of integration so that the undisturbed state with $n_0=1$ has $\delta |b|^2 =0$. The other point at which $\delta |b|^2 =0$ is at $n_0=\gamma$, while the positive maximum of $\delta |b|^2/\vA^2=\beta(\sqrt{\gamma}-1)^2$ occurs at $n_0=\sqrt{\gamma}$. Thus, for $\gamma<1$ ($\gamma>1$) there is a range of density decreases between $\gamma<n_0<1$ (density increases between $1<n_0<\gamma$) over which the magnetic field strength fluctuations are positive; they are negative outside these ranges. \markup{With $\gamma=0$, this equation is just the expression of pressure balance. As in the small-amplitude gyrokinetic case\cite{schektome2009}, pressure balance is violated only by the "finite-frequency" effects.}
\section{Steady nonlinear waves}
We now turn to studying steady nonlinear solutions of (\ref{eq:master}), i.e., solutions $n_0=n_0(\xi-\lambda\zeta)$, so that we may write $\ddze=-\lambda\ddxi$. \markup{In the plasma frame, the wave also moves at $\vA\cos\theta$ in $x$; $\lambda$ is is the small perturbation to the Alfv\'en velocity. To be precise, in the plasma frame, 
\beq
\ddt = -\left(\vA\cos\theta+ \lambda\beta\frac{\vA^2}{b_T^2} \vA\cos\theta\right) \ddx,
\eeq
so that in the case where $b_T=\vA$, $\lambda$ is the perturbed wave velocity in units of $\beta$ (which is small) times the Alfv\'en velocity $\vA$.} Then, we may integrate (\ref{eq:master}) once, obtaining
\begin{align}
\ddxi^2&\left[\log n_0 + \frac{\gamma}{2n_0^2}\right]\nonumber\\&=-\left[(n_0-1)^2-(n_0-1)\log n_0 - \gamma\left\{\frac{(n_0-1)^3}{2n_0^2}\right\}\right.\nonumber\\&\left.\phantom{\frac{(n_0-1)^3}{2n_0^2}}+\lambda (n_0-1)\right],
\end{align}
where we have chosen the constant of integration so that there is an equilibrium at $n_0=1$. We now change the dependent variable from $\xi$ to $n_0$ using
\beq
\frac{\mathrm{d}}{\mathrm{d}\xi} = \frac{\mathrm{d}n_0}{\mathrm{d}\xi} \frac{\mathrm{d}}{\mathrm{d}n_0},
\eeq
obtaining a differential equation for $[dn_0/d\xi]^2$,
\beq
\frac{1}{2}\frac{\mathrm{d}f}{\mathrm{d}n_0}\frac{\mathrm{d}}{\mathrm{d}n_0}\left(\left[\frac{\mathrm{d}n_0}{\mathrm{d}\xi}\right]^2\right) + \left[\frac{\mathrm{d}n_0}{\mathrm{d}\xi}\right]^2 \frac{\mathrm{d}^2f}{\mathrm{d}n_0^2}+  g,
\eeq
where
\begin{align}
f&=\log n_0 + \frac{\gamma}{2n_0^2}, \\
g&=(n_0-1)^2-(n_0-1)\log n_0 - \gamma\left\{\frac{(n_0-1)^3}{2n_0^2}\right\}+\lambda (n_0-1).
\end{align}
This has general solution
\beq
\left[\frac{\mathrm{d}n_0}{\mathrm{d}\xi}\right]^2=\left[\frac{\mathrm{d}f}{\mathrm{d}n_0}\right]^{-2}\left(D-2\int g\frac{\mathrm{d}f}{\mathrm{d}n_0}\mathrm{d}n_0\right)
\eeq
Noting that
\beq
\frac{\mathrm{d}f}{\mathrm{d}n_0} = \frac{n_0^2-\gamma}{n_0^3},
\eeq
we can anticipate that the solution becomes singular (i.e. forms a shock) at $n_0=\sqrt{\gamma}$. 
The integral is
\begin{align}
\int g \frac{\mathrm{d}f}{\mathrm{d}n_0}\mathrm{d}n_0 =\quad &\frac{1}{2}\left[(n_0-1)^2+\log^2n_0-2(n_0-1)\log n_0\right]\nonumber\\
+&\frac{\gamma}{2n_0^2}\left[(n_0-1)^2\log n_0-(n_0-1)^3\right]\nonumber\\
+&\frac{\gamma^2(n_0-1)^4}{8n_0^4}+\lambda[n_0-1-\log n_0]-\frac{\lambda\gamma(n_0-1)^2}{2n_0^2}.
\end{align}
We set the constant of integration so that at $n_0=1$, $\mathrm{d}n_0/\mathrm{d}\xi=0$: it will be adjusted by altering $D$ later. So, the solution is
\begin{align}
    &\left[\frac{\mathrm{d}n_0}{\mathrm{d}\xi}\right]^2\nonumber\\&=\frac{n_0^6}{(n_0^2-\gamma)^2}\left\{\frac{}{}D+\left[2(n_0-1)\log n_0 - \log^2 n_0 -(n_0-1)^2\right]\right.\nonumber\\
    &\quad+\frac{\gamma}{n_0^2}\left[(n_0-1)^3-(n_0-1)^2\log n_0\right]\nonumber\\
    &\quad\left.-\frac{\gamma^2(n_0-1)^4}{4n_0^4}+2\lambda\left[\log n_0+1-n_0\right]+\frac{\lambda\gamma(n_0-1)^2}{n_0^2}\right\}.
\end{align}
\markup{This is in the form of a Sagdeev pseudopotential\cite{sagdeev1962,chen1984}: similar techniques have been used to analyse Alfv\'en solitons in previous work \citep{hasegawa1976,shukla1982,wu1995,wu1996,berthomier1999,seyler1999,wu2003,wu2004}.} Solutions are confined to $[\mathrm{d}n_0/\mathrm{d}\xi]^2\geq0$. Starting at a value of $n_0$ for which $\mathrm{d}n_0/\mathrm{d}\xi=0$, the solution traces out a path in $\xi,n_0$ until another point at which $\mathrm{d}n_0/\mathrm{d}\xi=0$ is reached. $\dndxisq$ may in fact be written as
\begin{align}
    \left[\frac{\mathrm{d}n_0}{\mathrm{d}\xi}\right]^2=\frac{n_0^6}{(n_0^2-\gamma)^2}&\left\{\frac{}{}D+4p(\lambda-p)\right\},\label{eq:dndxi2}
\end{align}
where
\begin{figure*}[t]
    \centering
    \includegraphics[width=\linewidth]{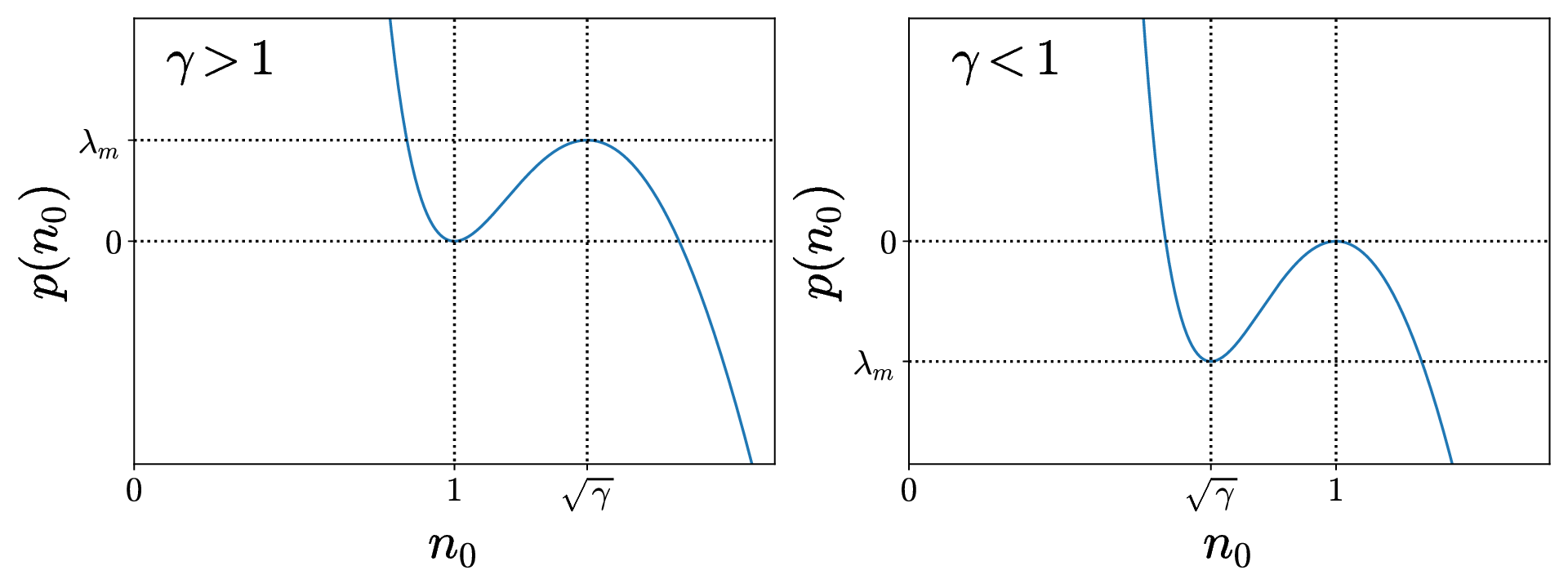}
    \caption{The function $p(n_0)$ in two cases with $\gamma>1$ (left) and $\gamma<0$ (right).}
    \label{fig:p}
\end{figure*}
\beq
p(n_0)=\frac{1}{2}\left[\frac{\gamma(n_0-1)^2}{2n_0^2}+\log n_0 -(n_0-1)\right].\label{eq:p}
\eeq
This function allows us to classify the possible steady waves, and is sketched in two representative cases in Figure~\ref{fig:p}. The behaviour is different depending on whether $\gamma>1$ ($\cos^2\theta>\beta$, "cold" waves) or whether $\gamma<1$ ($\cos^2\theta<\beta$, "warm" waves). The local extrema of $p(n_0)$ are 
\begin{align}
n_0=1&:\quad\text{min (max) with } p(1)=0\nonumber\\&\quad\quad\text{for }\gamma>1\text{ }(\gamma<1),\nonumber\\
n_0=\sqrt{\gamma}&: \quad\text{max (min) with } \lambda_m>0\text{ }(\lambda_m<0)\nonumber\\&\quad\quad\text{for }\gamma>1\text{ }(\gamma<1),
\end{align}
where $\lambda_m=p(\sqrt{\gamma})$ is the value at the local extremum. Importantly, $\dndxisq$ also has a singularity at $n_0=\sqrt{\gamma}$ due to the zero in the denominator there. This will cause some of the solutions to become singular, i.e. have formally discontinuous density profiles.
\subsection{Solitons}
\begin{figure*}[t]
    \centering
    \includegraphics[width=\linewidth]{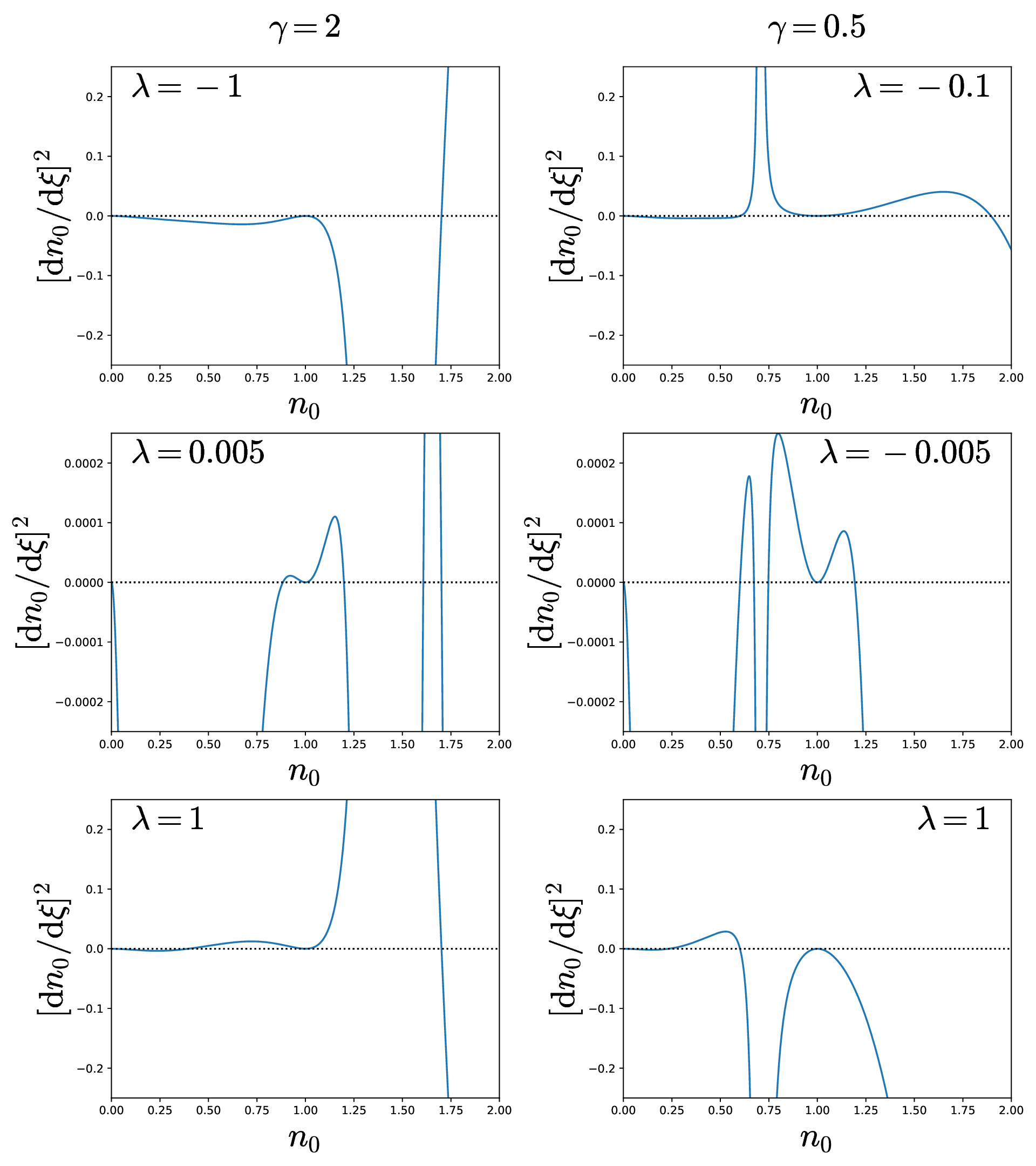}
    \caption{$\dndxisq$ for $\gamma>1$ (left) and $\gamma<1$ (right). All the cases identified in the text are shown: large negative velocity (top), large positive velocity (bottom), and small velocity between zero and $\lambda_m$ (middle).}
    \label{fig:dndxi2}
\end{figure*}
We have chosen the constant of integration so that $D=0$ corresponds to the solitary solutions. For $D=0$, the roots are at $n_0=0$, and when either $p(n_0)=0$ (attained when $n_0=1$ and $n_0=n_*$) or $p(n_0)=\lambda$ (attained at $n_0=n_{\lambda i}$). To reiterate our boundary conditions, we are in general allowing the magnetic field to point in different directions at $\pm\infty$: these solitary solutions thus correspond to "kinks" connecting two different regions. The density only varies where the magnetic field is rotating, see Eq.~(\ref{eq:n0phi}).

For a solitary wave to exist, $p(\lambda-p)$ must be positive between $n_0=1$ and another root. So, $p$ and $\lambda-p$ must have the same sign in this range. We will consider the cases $\gamma>1$ and $\gamma<1$ separately. Examples of $\dndxisq$ are shown in Figure~\ref{fig:dndxi2}: soliton solutions correspond to positive regions of the curve connected to $n_0=1$.

\subsubsection{Cool solitons, $\gamma>1$}

There are three ranges of interest for $\lambda$, which determines which different types of soliton are possible.

\paragraph{$\lambda\leq0$.---} At $n_0=1$, $p$ has a minimum at zero while $\lambda-p(1)$ is negative; thus, the zero of $\dndxisq$ at $n_0=1$ is a maximum and there are no solitary waves. \markup{This case is shown in Fig.~\ref{fig:dndxi2}, top left.}

\paragraph{$0<\lambda\leq\lambda_m$.---} In this range, there are three values $n_\lambda$ for which $p(n_\lambda)=\lambda$; they are arranged along the $n_0$ axis as $n_{\lambda1}<1<n_{\lambda2}<\sqrt{\gamma}<n_{\lambda3}$. In the range $n_{\lambda1}\leq n_0 \leq n_{\lambda2}$, both $p(n_0)\geq0$ and $\lambda-p(n_0)\geq0$, so there are both density spike and density dip solitons. In the range $n_{\lambda2}<n_0<n_{\lambda3}$, however, $p(n_0)>0$ but $\lambda-p(n_0)<0$, so this range cannot be accessed by solitary waves. \markup{This case is shown graphically in Fig.~\ref{fig:dndxi2}, middle left plot. There is a maximum of $\dndxisq$ around $n_0=1$, and the density spike and dip solitons correspond to excursions in opposite directions from this point up the the next crossing of $\dndxisq=0$. Note that the discontinuity in $\dndxisq$ at $\sqrt\gamma$ is at greater $n$ than the peak density in the soliton.} The density spike and dip solitons behave much like the mKdV solitons described in the previous section (and this correspondence becomes increasingly good as $\lambda\to0$), and involve a full rotation of the transverse magnetic field.

\begin{figure*}[t]
    \centering
    \includegraphics[width=0.9\linewidth]{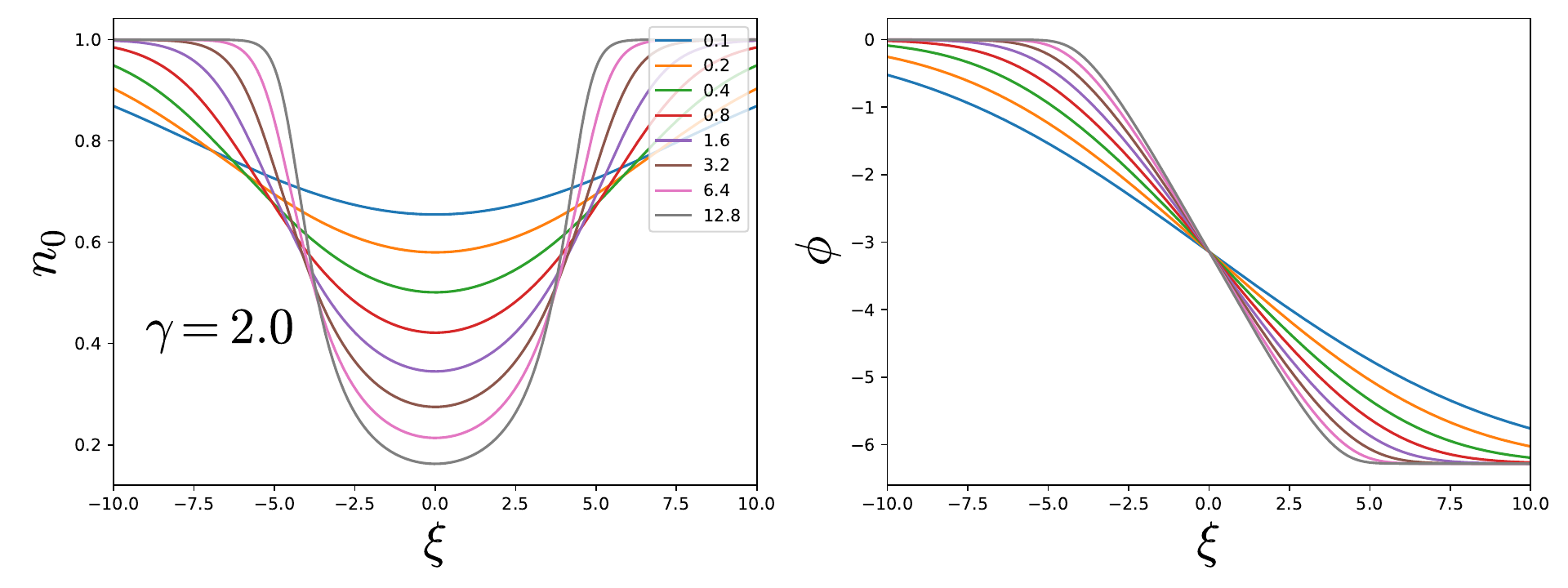} \\
    \includegraphics[width=0.9\linewidth]{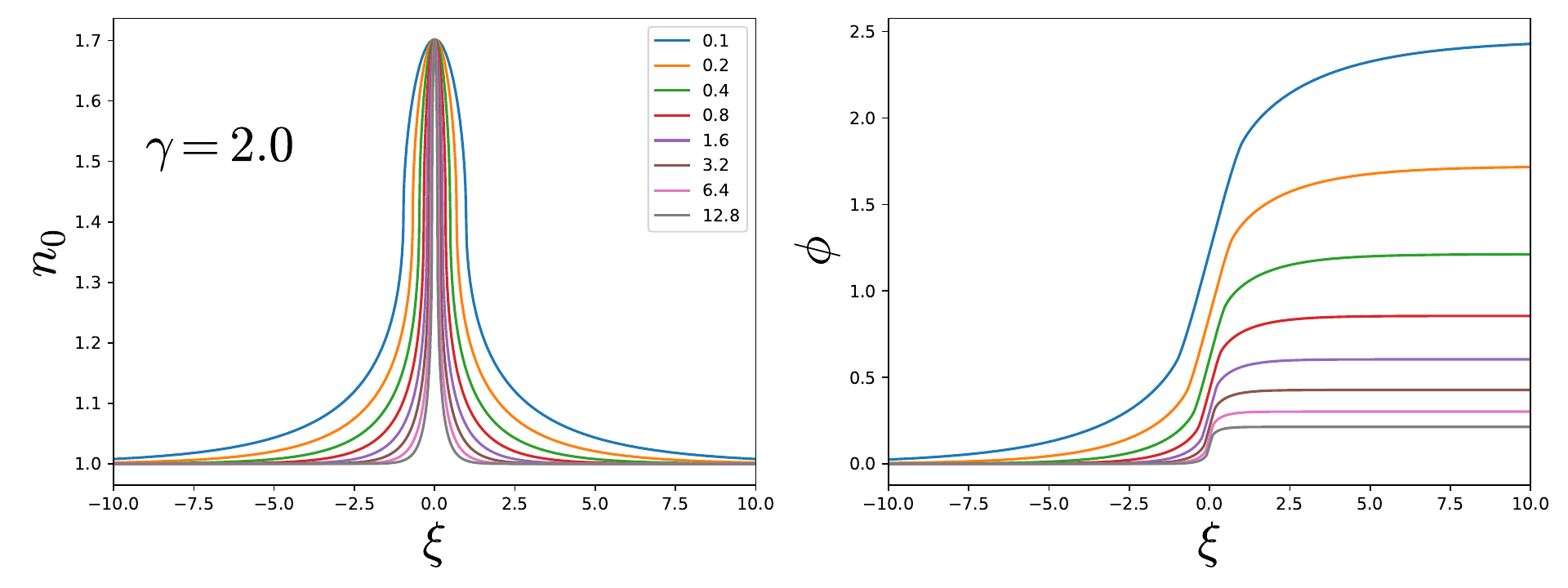}\\
    \includegraphics[width=0.9\linewidth]{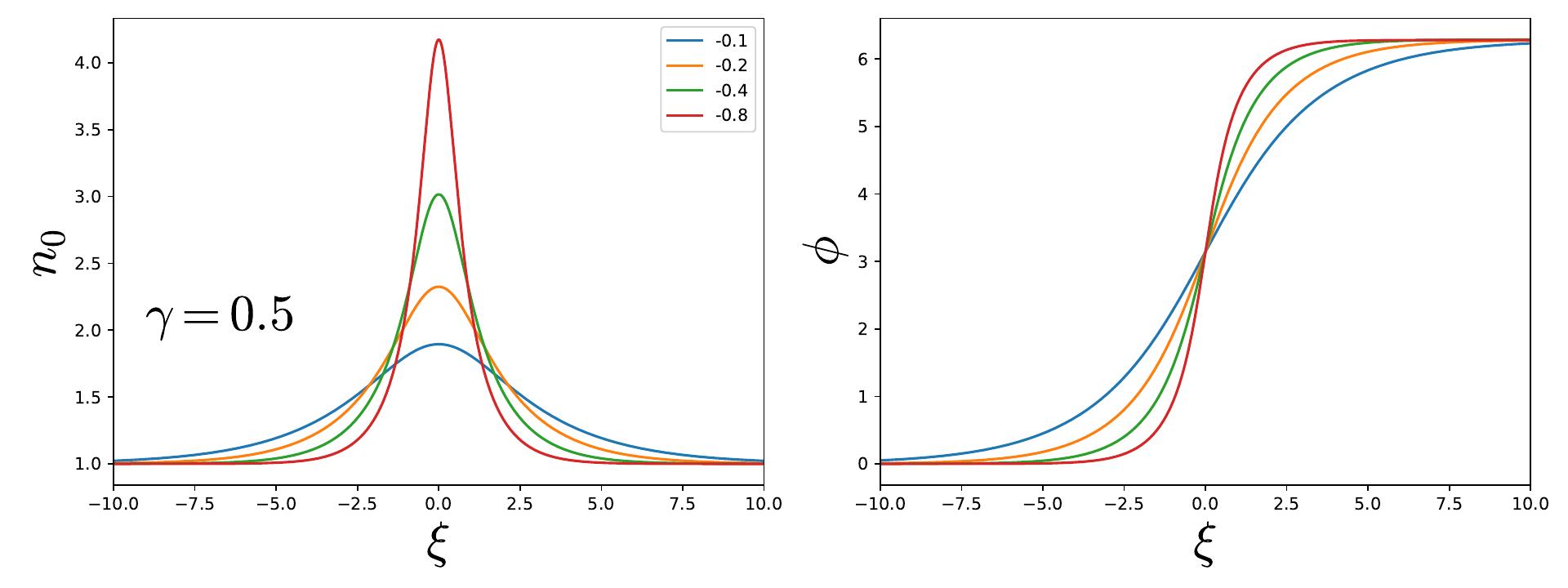}\\
    \includegraphics[width=0.9\linewidth]{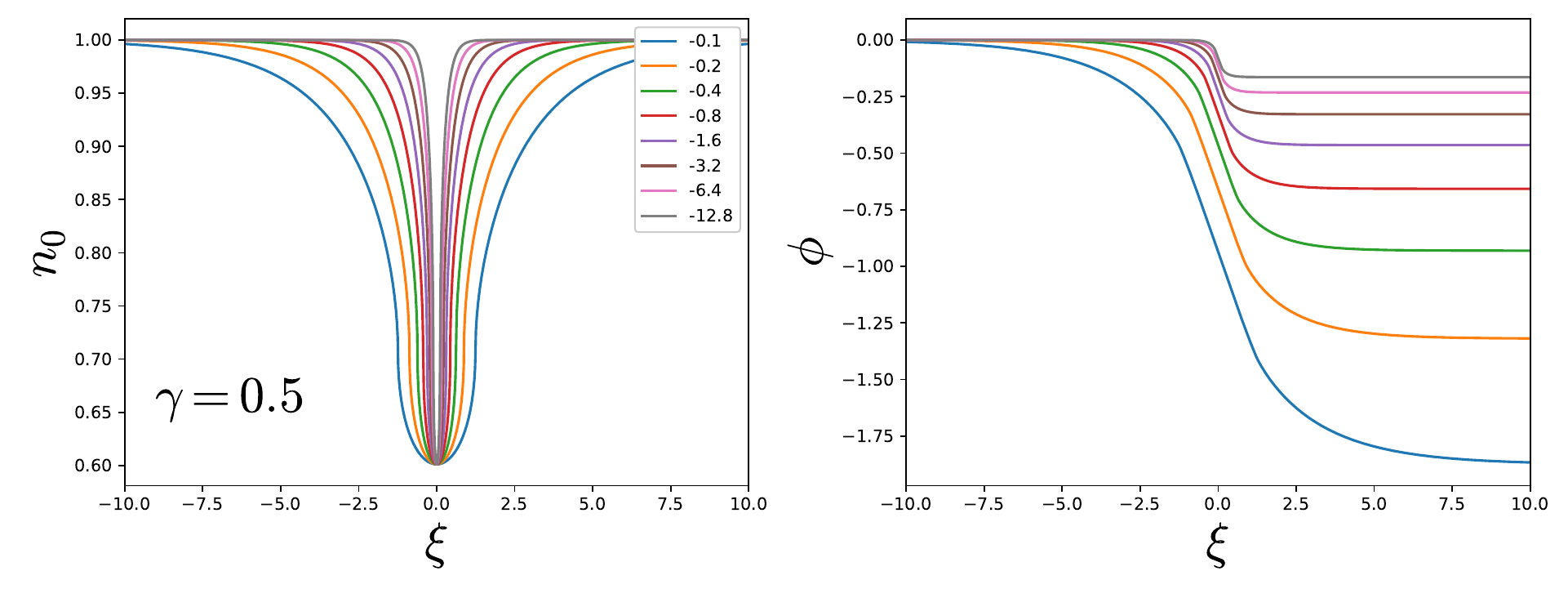}
    \caption{Left and right plots show the density and magnetic field rotation angle as a function of $\xi=x/d_i$ for the following cases (top to bottom rows): $\gamma>1$ dips, $\gamma>1$ (singular) spikes, $\gamma<1$ spikes, $\gamma>1$ (singular) dips. Different colors correspond to different values of $\lambda$, as labelled.}
    \label{fig:solwf}
\end{figure*}

\paragraph{$\lambda>\lambda_m$.---} Here, there is again only a single value of $n_\lambda$ solving $p(n_\lambda)=\lambda$, and $n_\lambda<1$. For $n_\lambda\leq n_0\leq1$, $p(n_0)\geq0$ and $\lambda - p(n_0)\geq 0$, so there is a density dip soliton. Graphically, $\dndxisq$ is shown for this case in the bottom left of Fig.~\ref{fig:dndxi2}, and corresponds to an excursion to lower density away from the minimum at $n_0=1$ to the next zero of $\dndxisq$ at $n_0<1$. This soliton involves a full rotation of the transverse magnetic field, and behaves similarly to the mKdV dip soliton. \markup{The waveforms for this case} are shown in the top row of Fig.~\ref{fig:solwf}. Note that because $n_0>0$, there is a maximum gradient of $\Phi$, attained as $\lambda\to\infty$. Because $\Phi$ rotates by $2\pi$, this also translates into a minimum soliton width.

Consider now the range $1\leq n_0 \leq n_*$, where $\gamma>n_*>\sqrt{\gamma}>1$ is the other root of $p(n_0)=0$; there, $p(n_0)\geq0$ and $\lambda-p(n_0)\geq0$, but because $n_*>\sqrt{\gamma}$, and $n_g=\sqrt{\gamma}$ is the location of the zero of the denominator of $\dndxisq$, this is a \emph{singular} spike soliton: the gradient of the density profile becomes infinite at $n_0=\sqrt{\gamma}$. Note that $n^*$ (the amplitude of the soliton) is independent of $\lambda$; this means that the maximum gradient in $\Phi$ is also a constant, and that as $\lambda$ increases, the width of the solution decreases, as does the total rotation in $\Phi$. \markup{Graphically, $\dndxisq$ is also shown for this case in the bottom left of Fig.~\ref{fig:dndxi2}: an excursion from $n_0=1$ to zero of $\dndxisq$ at $n_0>1$ necessarily encounters the discontinuity at $n_0=\sqrt\gamma$. Integrating across this discontinuity results in a finite, discontinuous jump in $n_0$ in the waveforms on either side of the central spike: example waveforms} are shown in the second row of Fig.~\ref{fig:solwf}. \markup{Part of the rise and fall from the peak of each density waveform is thus in fact a discontinuous shock: this is what gives the curves the slightly ``bullet"-like shape, rather than the smoother appearance of the regular solitons.}
\subsubsection{Warm solitons, $\gamma<1$}
\begin{figure*}
\includegraphics[width=\linewidth]{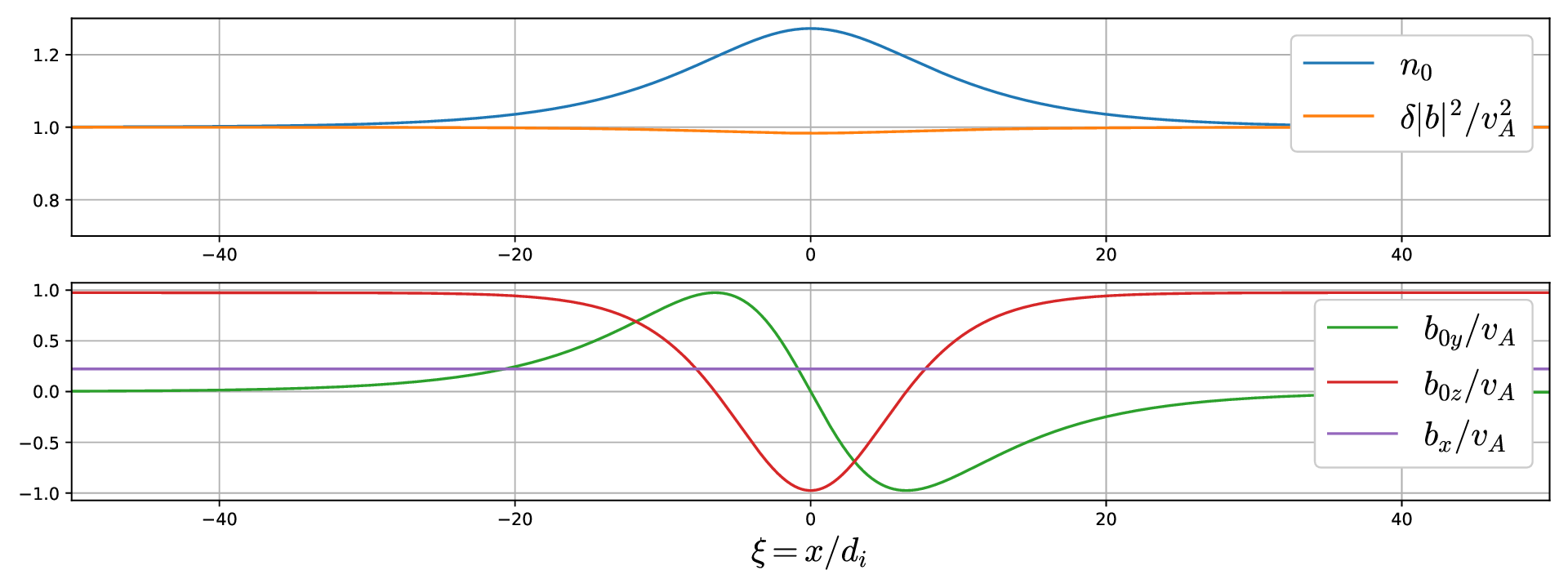}
\caption{The waveforms of $n_0$ (blue), $\delta |b|^2$ (orange), $b_{0y}$ (green), $b_{0z}$ (red), and $b_x$ (purple, constant) plotted as a function of $x$, as would be observed by a spacecraft whose velocity relative to the plasma was in the $x$-direction. For this plot, $\lambda=-0.01$, $\cos^2\theta=0.05$, $\beta=0.1$, and $b_T=\vA\sin\theta$.\label{fig:timeseries}}
\end{figure*}
The analysis in this case is entirely analogous to the case $\gamma>1$, so we will just summarise the results. For $\lambda\geq0$, there are no solitary waves \markup{($\dndxisq$ for this case is shown in the bottom right of Fig.~\ref{fig:dndxi2})}. For $\lambda_m\leq \lambda <0$, there are both spike and dip solitons, which involve full rotations of the transverse magnetic field and are similar to the mKdV solitons. \markup{$\dndxisq$ for this case is shown graphically in the middle right of Fig.~\ref{fig:dndxi2}: the discontinuity at $\sqrt{\gamma}$ is always at smaller $n_0$ than the maximum amplitude of the dip soliton.} For $\lambda<\lambda_m$, there are again regular density spike solitons which involve full rotations of the transverse magnetic field and are similar to the mKdV solitons\markup{: graphically, $\dndxisq$ is shown for this case in the top right panel of Fig.~\ref{fig:dndxi2} and corresponds to moving from $n_0=1$ to the zero at $n_0>1$. The waveforms for} this case are shown in the third row of Fig.~\ref{fig:solwf}. The width of these solutions decreases with $\lambda$, meaning the magnetic field rotation can get increasingly steep. This would break down when the amplitude of the density invalidates the ordering in (\ref{eq:constb}); at this point the large local density means that the local value of $\beta$ is order unity. A synthetic time-series plot that would be observed a spacecraft whose velocity relative to the plasma was exactly in the $x$-direction is shown in Fig.~\ref{fig:timeseries}.

There are also singular dip solitons for $\lambda<\lambda_m$, which involve infinite gradients in the density profile at $n_g=\sqrt{\gamma}$, minimum density $n_*$ independent of $\lambda$, and involve progressively smaller total rotations in $\Phi$ as $|\lambda|$ decreases. \markup{Graphically, $\dndxisq$ is shown in the top right plot of Fig.~\ref{fig:dndxi2}: a path from $n_0=1$ to the zero of $\dndxisq$ at $n_0<1$ necessarily crosses the discontinuity at $n_0=\sqrt{\gamma}$. Integrating across this discontinuity results again in a finite, discontinuous jump in $n_0$ on either side of the central density dip: example waveforms} are shown in the bottom row of Fig.~\ref{fig:solwf}.

\subsubsection{Magnetic field strength behaviour in the solitons}
The behaviour of the magnetic field strength can be deduced from Eq.~(\ref{eq:magfieldstrength}). For "cool solitons" with $\gamma>1$, density dips also have magnetic-field-strength dips, while density spikes (both singular and non-singular mKdV type solitons) have magnetic-field-strength spikes; this is guaranteed because $1>n^*<\gamma$. For "warm solitons" with $\gamma<1$, density spikes have magnetic field strength dips, while density dips (both singular and non-singular) have magnetic-field-strength spikes.

\markup{As mentioned in the introduction, NASA's PSP spacecraft has observed large-amplitude Alfv\'enic structures, dubbed ``switchbacks", in the near-Sun solar wind and corona: often, these structures have remarkably sharp, discontinuous boundaries. These boundaries systematically have a small, localised dip in the magnetic-field-strength $|\vB|$ -- see, for example, Figure 7 of Farrell et al. 2020\cite{farrell2020}. Superficially, this seems at odds with the analytic considerations above as well as the mKdV solitons, which can in principle have either sign of $|\vB|$ fluctuation. However, this symmetry disappears when one considers the case of a soliton that is sufficiently narrow compared to $d_i$, i.e. for $\lambda$ sufficiently large. Then, some of the solitons are of the singular type, and thus physically unrealizable: these occur when $\lambda>\lambda_m$ for $\gamma>1$ or $\lambda<\lambda_m$ for $\gamma<1$. The solitons that remain non-singular as the soliton width becomes small are the density dip solitons for $\gamma>1$ and the density spike solitons with $\gamma<1$: both of these have magnetic-field-strength dips. This may explain the asymmetry observed in the magnetic-field-strength fluctuations in the observed switchback edges, although a more complex model is needed to explain other features.}

\subsection{Periodic waves}
\begin{figure*}[t]
    \centering
    \includegraphics[width=\linewidth]{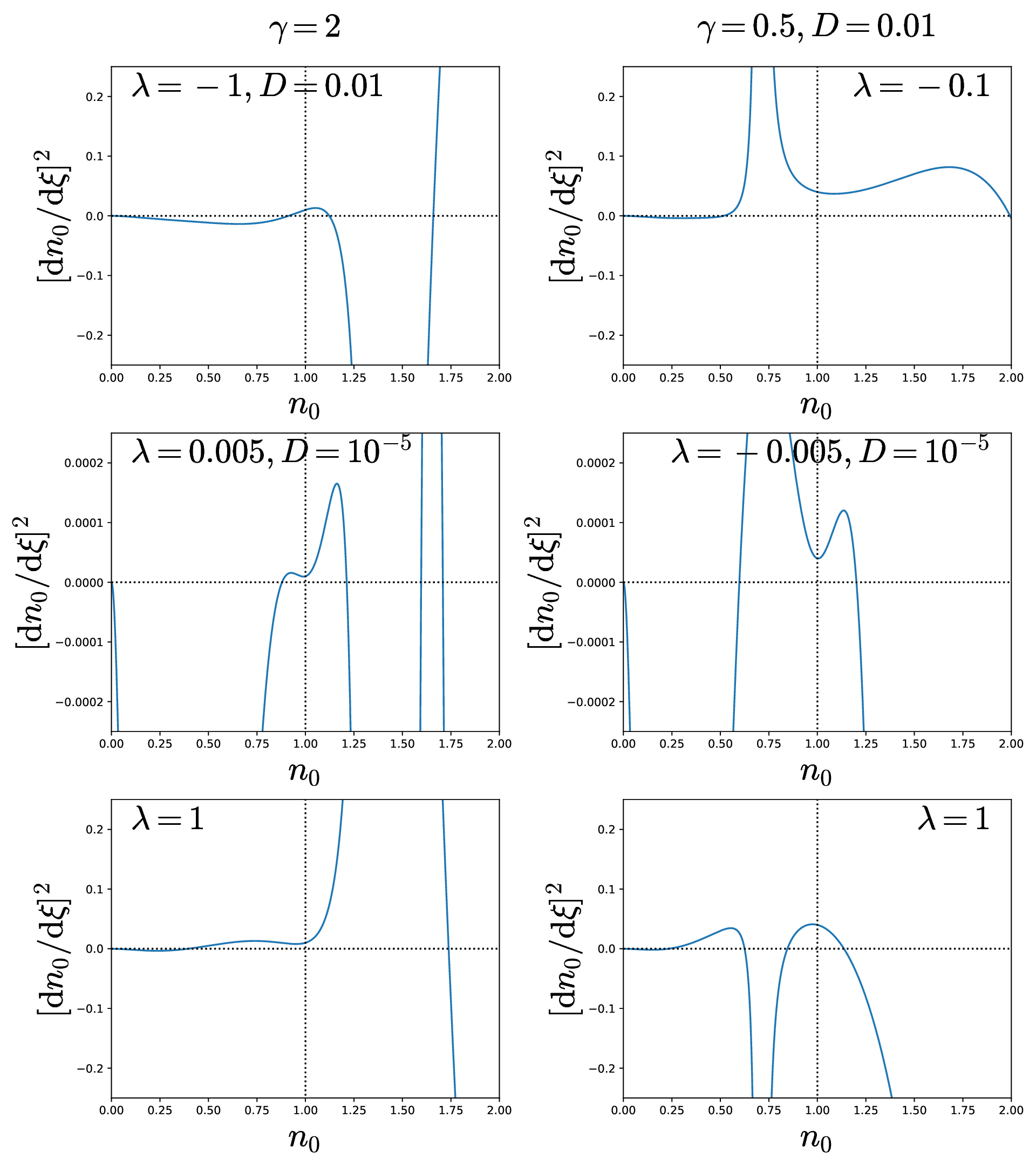}
    \caption{$\dndxisq$ for $\gamma>1$ (left) and $\gamma<1$ (right), with $D>0$. Continuous paths between zeros of $\dndxisq=0$ for which $\dndxisq\geq0$ correspond to the periodic waves. All the cases identified in the text are shown: large negative velocity (top), large positive velocity (bottom), and small velocity between zero and $\lambda_m$ (middle).}
    \label{fig:dndxi2pw}
\end{figure*}
For periodic waves, all that is needed is that there be some region of $\dndxisq>0$ around $n_0=1$; i.e., we may set $D\neq0$ in Eq.~\ref{eq:dndxi2}. The points at which the extreme values of the density may be found by solving for $\dndxisq=0$, and occur when $p=p^\pm$, with
\beq
p^\pm = \frac{\lambda \pm \sqrt{\lambda^2 + D}}{2}.
\eeq
The quantity in curly brackets in (\ref{eq:dndxi2}) is then
\beq
-(p-p^+)(p-p^-),
\eeq
and this must be non-negative in a range of $n_0$ containing $n_0=1$ for there to be a possible nonlinear wave. Note that $p^+>0$ and $p^-<0$, irrespective of the sign of $\lambda$, since $D>0$ for $\dndxisq>0$ at $n_0=1$. For $\lambda>0$, $p^+>\lambda$, while for $\lambda<0$, $p^-<\lambda$.

The physical behaviour of the waves can be understood roughly by examining the plots in Figure~\ref{fig:dndxi2}, shifting the relevant curves up slightly: this is shown in Figure~\ref{fig:dndxi2pw}. The periodic waves come in two fundamental types: one exists in the small-amplitude limit, while the other consists of repeated pairs of solitons (one dip, one spike), and thus does not exist at small amplitude.
\subsubsection{Cool waves, $\gamma>1$}
\paragraph{$\lambda>0$.---} 

For $p^+<\lambda_m$, $p(n_0)=p^+$ at three values of $n_0$, $n_{+1}<1<n_{+2}<\sqrt{\lambda}<n_{+3}$. Between all of these, $p-p^->0$, but only between $n_{+1}$ and $n_{+2}$ is $p-p^+<0$, so that $\dndxisq>0$. This means that the range between $n_{+2}$ and $n_{+3}$ is inaccessible, and thus also so is $n_->n_{+3}$, the solution to $p(n_-)=p^-$. Thus, a solution moves between $n_{+1}$ and $n_{+2}$. It may be seen \markup{in the middle left plot of Fig.~\ref{fig:dndxi2pw} that for small $D$ the solution consists of repeated alternating dips and spikes; as $D\to 0$ the distance between the pairs of dips and spikes tends to infinity.}

For $p^+>\lambda_m$, $p(n_0)=p^+$ at only one value of $n_0$,$n_+<1$. Between $n_+$ and $n_-$, $p-p^+<0$ but $p-p^->0$, so a solution exists; however, because $n_->\sqrt{\gamma}$, and there is a zero in the denominator of $\dndxisq$ at $n_0=\sqrt{\gamma}$, this solution is singular. \markup{As an example, by slightly raising the curve in the middle left-hand panel of Figure~\ref{fig:dndxi2}($\gamma=2$, $\lambda=1$) by choosing $D>0$, to make the curve in the middle left plot of Figure~\ref{fig:dndxi2pw}, one can see that for small $D$ the solution consists of repeated alternating dips and singular spikes; as $D\to0$ the distance between them tends to infinity.}

\paragraph{$\lambda<0$.---}

In this case, exactly the same analysis of the roots applies: however, note that now it is $p^-<\lambda$, while for $D\to 0$ $p^+\to0$. This means that the first two roots of $p(n_0)=p^+$, $n_{+1}$ and $n_{+2}$, approach $n_0=1$ from below and above respectively. Thus this solution connects to the small-amplitude periodic waves. \markup{This may also be seen by slightly raising the curve in the top left-hand panel of Figure~\ref{fig:dndxi2} ($\gamma=2,\lambda=-1$) by choosing $D>0$, to make the curve in the top left-hand panel of Figure~\ref{fig:dndxi2pw}: a small region of positive $\dndxisq$ appears around $n_0=1$.}

For large enough $D$, $p^+>\lambda_m$, and only a solution involving a discontinuity exists, between $n_+$ and $n_-$.

As expected from the dispersion relation (\ref{eq:slightkawdisp}), this branch, connected to the small-amplitude waves, (with $\gamma=\cos^2\theta/\beta>1$) has $\lambda<0$. It is interesting that the opposite is true for the soliton-like solutions.

\subsubsection{Warm waves, $\gamma<1$}

An analogous argument can be applied to the waves with $\gamma<1$: for $\lambda<0$, the periodic waves behave like pairs of regular dip and regular spike solitons for $p^->\lambda_m$, and pairs of singular dip and regular spikes for $p^-<\lambda_m$. For $\lambda>0$, one obtains nonlinear corrections to the small-amplitude waves, which contain singularities for sufficiently large $D$ such that $p^-<\lambda_m$. It is natural from the linear dispersion relation (\ref{eq:slightkawdisp}) that such waves have $\lambda>0$. The three types of periodic waves with $\gamma<1$ are shown in the right-hand panels of Fig.~\ref{fig:dndxi2pw}.

\begin{table}
\centering
        \begin{tabular}{|l|l|l|}
            \hline
            \multicolumn{1}{|c}{$\gamma$} & \multicolumn{1}{c}{dips} & \multicolumn{1}{c|}{spikes} \\ \hline
            \multirow{2}{*}{$>1$} & regular, $\lambda>0$ & regular, $\lambda_m>\lambda>0$ \\
                                 &  & singular, $\lambda>\lambda_m$ \\\hline
            \multirow{2}{*}{<1} & regular, $\lambda_m <\lambda<0$ & regular, $\lambda<0$ \\
                                 & singular, $\lambda<\lambda_m$& \\
            \hline
        \end{tabular}
\caption{Summary of the soliton regimes identified in the paper; $\gamma = \cos^2\theta / \beta$ and $\lambda_m = p(\sqrt{\gamma})$ (see Eq. \ref{eq:p}).\label{tab:summary}}
\end{table}

\section{Discussion}\label{sec:disc}
We have shown that for one-dimensional, large-amplitude AW with perpendicular lengthscales comparable to $d_i$, non-trivial nonlinear evolution of the AW is \textit{entirely enabled by the dispersive terms}: at zeroth order in our expansion, we just have a constant-magnetic-field-strength solution, similarly to MHD, but with large density fluctuations, localised where there are magnetic field rotations. It is only at first order that non-trivial evolution of the waves appears, dependent on the dispersive parameters $k^2\rho_s^2$ and $\kpar^2 d_i^2$. 

Second, in an extension of the theory resulting in the modified Korteweg-de Vries (mKdV) equation\cite{kakutani1969,kawahara1969}, we have shown that non-singular AW solitons still exist even when $d_i\ddx\sim 1$, taking the form of full rotations of the transverse magnetic field: these solutions cannot be found if one takes a small-amplitude approximation, even with strong nonlinearity\cite{seyler1999,mallet2023}. If the structure has a large width compared to $d_i$, our dynamical equation reduces to the mKdV equation, with both dip and spike solitons. However, for sufficiently small-scale structures, and depending on the parameter $\gamma = \cos^2\theta/\beta$, where $\theta$ is the angle of propagation relative to the background magnetic field and $\beta=c_s^2/\vA^2$, only density dip ($\gamma>1$) or density spike ($\gamma<1$) non-singular solitons are possible. The different soliton regimes are summarized in Table~\ref{tab:summary}. We also categorize the periodic nonlinear waves, which appear as two types: the first correspond to nonlinear corrections to the standard (linearized) small-amplitude waves, while the second corresponds to repeated pairs of large-amplitude solitons. Alongside the non-singular large-amplitude solitons, we have also (re-)discovered families of singular dip and spike solitons, which involve a pair of discontinuities in the density profile; similarly to the small-amplitude case\citep{seyler1999}. Unlike the non-singular solitons, these are characterized by a fixed maximum density drop or increase, and a magnetic-field rotation that can be small: the narrower the soliton in $\xi = x/d_i$, the smaller the magnetic-field rotation. In the real world, such discontinuous solutions would no doubt be strongly dissipative: the fluid model used here is grossly insufficient, and thus we expect these solutions to be modified significantly in a more realistic model.

These results provide an interesting extension to the work of Seyler \& Lysak (1999)\citep{seyler1999}, and may be relevant to the physics of the nearly-discontinuous Alfv\'enic switchback edges observed recently by PSP. In particular, it may explain why some switchback edges appear to be rather stable structures, persisting for relatively long times. Observationally, switchbacks often seem to have a magnetic-field-strength dip (or "dropout") at their boundaries\cite{farrell2020}. We show that this follows naturally from our analysis: only solitons with magnetic field strength dips can survive without developing singularities as the magnetic-field rotation becomes sufficiently steep. However, it should be noted that many features of the switchback edges are not explained by the current model: for example, the only solitons in this model require a full rotation of the transverse magnetic field, while switchbacks exist with a wide range of amplitudes. Again, this may be due to our neglect of kinetic processes such as Landau damping \markup{as well as finite ion temperature}: such effects have been the subject of past work in the quasi-parallel case\citep{mjolhus1988,medvedev1996}, and we plan to investigate this in future work.

\acknowledgements
AM is grateful to S. Dorfman, T. Bowen, M. Abler, C. Chen, C. Chaston, J. Bonnell, S. Boldyrev, and J. Squire for useful discussions, and was supported by NASA grant 80NSSC21K0462 and NASA contract NNN06AA01C.

\bibliography{solitons}

\end{document}